\documentstyle[12pt,epsf]{article}
\newcommand{\bec}{\begin{center}}
\newcommand{\ec}{\end{center}}
\newcommand{\bee}{\begin{equation}}
\newcommand{\ee}{\end{equation}}
\newfont{\blackboard}{msbm10 scaled\magstep2}
\newcommand{\Z}{\mbox{\blackboard\symbol{"5A}}}

\begin{document}
\large
\begin{titlepage}
\bec
{\Large\bf  K-Theory and Gauge Solitons \\}
\vspace*{15mm}
{\bf Yu. Malyuta \\}
\vspace*{10mm}
{\it Institute for Nuclear Research\\
National Academy of
Sciences of Ukraine\\
252022 Kiev, Ukraine\\}
e-mail: interdep@kinr.kiev.ua\\
\vspace*{35mm}
{\bf Abstract\\}
\ec
The spectrum of $D$-brane charges
in the Type IIB string theory
compactified on $AdS_{p+2}\times S^{8-p}$
is computed using the K-theory
approach. The  result differs from the
corresponding result presented in 
the digest of Olsen and Szabo.
\end{titlepage}
\section{Introduction}
$D$-branes play a significant role in 
supersymmetric string and field theories.
Two of the most outstanding 
developments in this direction have been
achieved:\\
1. The generalized AdS/CFT duality 
\cite{1.}, which relates the 
superconformal field theory on
$Dp$-branes placed at the orbifold 
singularity and the Type IIB string
theory compactified on 
$AdS_{p+2}\times H^{8-p}$\ ;\\
2. The K-theory approach to $D$-brane
charges \cite{2.,3.}, which identifies 
$D$-brane charges with elements
of Grothendieck K-groups \cite{4.}
of horizon manifolds $H^{8-p}$\ .

	In the present paper we use
K-theory to compute the spectrum of
$D$-brane charges in the Type IIB
string theory compactified on 
$AdS_{p+2}\times S^{8-p}$. 
The  result differs from the
corresponding result presented in
\cite{5.}.

\section{$D$-brane charges}
Let us equip the horizon $S^{8-p}$
with the gauge bundle

\bec
\begin{tabular}{ccc}
$U$\hspace*{-0.2cm}&$\rightarrow$ &
\hspace*{-0.8cm}$E$ \\
&    &\hspace*{-0.8cm}$\downarrow$ \\
 &   &\hspace*{-3mm}$S^{8-p}$  \\
\end{tabular}
\ec
The Steenrod classification 
theorem \cite{6.} asserts that
this bundle is characterized 
by the homotopy group
$\pi_{7-p}(U)$.
Using  the standard definition
of the K-group \cite{7.}, we obtain
\bee
\widetilde{K}(S^{8-p})=\pi_{7-p}(U) \ .
\ee
The exact homotopy sequence \cite{7.}

\[\hspace*{-1cm}\ldots \rightarrow 
\pi_{n}\biggl(U(2N)/U(N)\biggr)
\rightarrow \pi_{n}\biggl(U(2N)/U(N) 
\times U(N)\biggr) \rightarrow \]
\[\hspace*{1cm}\rightarrow 
\pi_{n-1}\biggl(U(N)\biggr) 
\rightarrow \pi_{n-1}\biggl(U(2N)/U(N)\biggr)
\rightarrow \ldots \]
\vspace*{-5mm}\\
for the universal bundle
\vspace*{5mm}
\bec
\begin{tabular}{ccc}
$U(N)$\hspace*{-0.2cm}&$\rightarrow$ &
\hspace*{-1.8cm}$U(2N)/U(N)$ \\
&    &\hspace*{-1.7cm}$\downarrow$ \\
 &   &\hspace*{-2mm}$U(2N)/U(N)\times U(N)$  \\
\end{tabular}
\ec
\vspace*{3mm}
yields

\[\widetilde{K}(S^{8-p})=
\pi_{7-p}(U)=\pi_{8-p}(B_{U}) \ ,\]

\hspace*{-6mm}where $B_{U}$ is the inductive 
limit of the manifold
\bee
U(2N)/U(N)\times U(N) \ .
\ee

	The manifold (2) has the following 
interpretation in terms of $D$-branes \cite{8.}.
When $2N$ coinciding branes are separated
to form two parallel stacks of $N$ coinciding
branes, their gauge symmetry $U(2N)$ is
spontaneously broken to $U(N)\times U(N)$.
This situation generically allows for the
existence of gauge solitons. 

	$D$-brane charges take values in the
K-groups (1). To compute K-groups (1), we 
use the Bott periodicity theorem \cite{9.}.
The spectrum of $D$-brane charges is
recorded in Table 1.

\bec

Table 1\\
\vspace*{7mm}
\begin{tabular}{|c|c|c|c|c|c|c|c|c|c|c|c|} 
\hline
\normalsize$Dp$ &\normalsize$D8$ 
&\normalsize$D7$ &\normalsize$D6$ 
&\normalsize$D5$ &\normalsize$D4$ 
&\normalsize$D3$ &\normalsize$D2$ 
&\normalsize$D1$ &\normalsize$D0$ 
&\normalsize$D(-1)$ &\normalsize$D(-2)$  
\\  \hline
\normalsize$S^{n}$&\normalsize$S^{0}$ 
&\normalsize$S^{1}$ &\normalsize$S^{2}$ 
&\normalsize$S^{3}$ &\normalsize$S^{4}$ 
&\normalsize$S^{5}$ &\normalsize$S^{6}$ 
&\normalsize$S^{7}$ &\normalsize$S^{8}$ 
&\normalsize$S^{9}$ &\normalsize$S^{10}$  
\\ \hline
\normalsize$\widetilde{K}(S^{n})$ 
&\normalsize$\Z$ &\normalsize0 
&\normalsize$\Z$ &\normalsize0 
&\normalsize$\Z$ &\normalsize0 
&\normalsize$\Z$ &\normalsize0 
&\normalsize$\Z$ &\normalsize0 
&\normalsize$\Z$  \\ \hline
\end{tabular}\\   
\ec
\vspace*{5mm}
\section{Remark}
The result recorded in Table 1 
differs from the corresponding
result obtained in \cite{5.}:
\vspace*{7mm}
\bec

\begin{tabular}{|c|c|c|c|c|c|c|c|c|c|c|c|}
\hline
\normalsize$Dp$ &\normalsize$D9$
&\normalsize$D8$ &\normalsize$D7$
&\normalsize$D6$ &\normalsize$D5$
&\normalsize$D4$ &\normalsize$D3$
&\normalsize$D2$ &\normalsize$D1$
&\normalsize$D0$ &\normalsize$D(-1)$  
\\  \hline
\normalsize$S^{n}$&\normalsize$S^{0}$
&\normalsize$S^{1}$ &\normalsize$S^{2}$
&\normalsize$S^{3}$ &\normalsize$S^{4}$
&\normalsize$S^{5}$ &\normalsize$S^{6}$
&\normalsize$S^{7}$ &\normalsize$S^{8}$
&\normalsize$S^{9}$ &\normalsize$S^{10}$  
\\ \hline
\normalsize$\widetilde{K}(S^{n})$
&\normalsize$\Z$ &\normalsize0
&\normalsize$\Z$ &\normalsize0
&\normalsize$\Z$ &\normalsize0
&\normalsize$\Z$ &\normalsize0
&\normalsize$\Z$ &\normalsize0
&\normalsize$\Z$  \\ \hline
\end{tabular}\\
\ec

\end{document}